\documentclass[aps,prl,amsmath,amssymb,preprintnumbers,nofootinbib,twocolumn,showpacs,floatfix]{revtex4}
\usepackage{epsfig,graphics,color}
\usepackage{graphicx}
\usepackage{dcolumn}
\usepackage{bm}
\usepackage{amsmath}

\newcommand \ra  {\rightarrow}

\newcommand \A {\alpha}

\newcommand \prt {\partial}

\newcommand \bvec{\left( \begin{array}{c} }
\newcommand \evec{\end{array} \right)}

\newcommand \bea{\begin{eqnarray} }
\newcommand \eea{\end{eqnarray} }
\newcommand \nn {\nonumber}
\newcommand {\be} {\begin{equation}}
\newcommand {\ee} {\end{equation}}

\newcommand {\mbx} {\mbox{}}

\newcommand {\ata} {& \times &}

\begin{document}

\title{The in-medium scale evolution in jet modification} 

\author{A.~Majumder}
\affiliation{Department of Physics, Duke University, Durham, NC 27708, USA}
\affiliation{Department of Physics, The Ohio State University, Columbus, OH 43210, USA}

\date{ \today}

\begin{abstract}
The in-medium modification of the scale dependence of the fragmentation function in dense matter, brought 
about by higher twist corrections to the Dokshitzer-Gribov-Lipatov-Altarelli-Parisi (DGLAP) 
evolution equations, is derived.  
A phenomenologically motivated resummation is outlined which incorporates the next-to-leading twist single 
gluon emission kernel along with the vacuum emission kernel and provides an in-medium 
virtuality evolution of the final fragmentation function of a hard jet propagating through 
dense matter. The concept of a fragmentation function is generalized to include a dependence on distance travelled in the 
medium. Following this, numerical implementations are carried out and compared to 
experimental results on the single inclusive suppression observed in Deep-Inelastic 
scattering (DIS) off a large nucleus. 
\end{abstract}


\maketitle




With the commissioning of the Large Hadron Collider (LHC) heavy-ion experiments, the study of 
hard probes in deconfined matter will have entered the deep 
perturbative domain: one expects jets with energies of several hundred GeV to be 
produced, plough through the dense deconfined matter and fragment far outside the medium~\cite{Abreu:2007kv}. 
With the start of the future Electron-Ion 
Collider (EIC), a similar step will be taken in the study of cold nuclear 
matter with hard jets. Current experiments at the Relativistic Heavy-Ion Collider (RHIC)~\cite{RHIC_Whitepapers,highpt} and 
at DESY (HERMES)~\cite{Airapetian:2000ks} have just begun to probe the perturbative regime. At RHIC 
this is believed to start above a rather high transverse momenta $p_T \sim 6-7$ GeV (due to the 
expanding medium)~\cite{Bass:2008rv}. In DIS, the extent of the perturbative region depends on the 
energy ($\nu$) lost by
the electron, the fraction carried by the detected hadron and the size of the 
struck nucleus~\cite{Majumder:2004pt}.

In the case of single-hadron-inclusive suppression in DIS on a large nucleus $A$, 
the application of perturbative methods is based on 
the factorized cross section to produce a hadron with a fraction $z$ of the photon light-cone  momentum $q^-$~\cite{col89}:
\bea
\frac{d \sigma^h (Q^2)}{d z} = \sum_{i}F_i (x_i, Q^2) \sigma_{i \ra j} (Q^2) \widetilde{D}_j^h(z,Q^2).  \label{pert_fact}
\eea
In this equation, $F_i(x_i)$ is the parton distribution function for a parton $i$ with momentum fraction $x_i$, 
$\sigma_{i \ra j}$ is the hard cross section with the virtual photon to produce 
parton $j$ and $\widetilde{D}$ is the medium modified fragmentation function to fragment into the hadron $h$ 
after undergoing multiple scattering in the nuclear medium~\cite{guowang}.

The multiple scattering of the hard parton in the medium generically has two parts, a perturbatively calculated 
contribution which represents the scattering and gluon radiation from the hard parton, and the non-perturbative 
distribution of the soft gluons off which the hard partons will scatter~\cite{HT_fact}. Both these parts in 
combination with the 
standard vacuum shower of gluons leading to the non-perturbative fragmentation into observable hadrons are 
included within the definition of the medium modified fragmentation function. 
The three different factorized parts in Eq.~\eqref{pert_fact} will necessarily depend on the 
factorization scales ($\mu_i$ for the initial state and 
$\mu_f$ for the final) which in Eq.~\eqref{pert_fact} have been chosen as the hard scale $Q^2$.   
So defined, the medium modified fragmentation functions will reduce to the standard vacuum fragmentation 
functions $D(z,Q^2)$, as the extent of the medium in the final state is reduced.

As the energies of the process are increased, all three factorized functions in Eq.~\eqref{pert_fact} change with $Q^2$. 
While volumes of work exist on the perturbatively calculable scale dependence of both the hard part 
and the nuclear structure functions (as well as vacuum structure and fragmentation functions)~\cite{AP,evol,Eskola:1998df}, 
little attention has been paid to the scale dependence of the medium modification of the fragmentation functions. 
In the case of DIS on a nucleon  or in $p$-$p$ collisions, i.e., in the absence of 
an extended medium, the vacuum fragmentation functions have a well known and perturbatively calculable 
dependence on the scale $Q^2$ of the process given by the DGLAP evolution equations~\cite{AP}: 
\bea
\mbx \!\!\! \frac{\prt D_q^h (z, Q^2)}{ \prt \log(Q^2) }\!\! &=&\!\! \frac{\A_s(Q^2)}{2\pi} \int_z^1 \frac{dy}{y} 
P_{q \ra i} (y) D_i^h \left( \frac{z}{y}, Q^2 \right) . \label{vac_DGLAP}
\eea
In the above equation, $\A$ is the strong coupling constant and $P_{q \ra i}$ represents the probability for a parton $q$ to 
split into an parton $i$ which carries a fraction $y$ of the original momentum and finally fragments into the hadron $h$ (a 
sum over $i$ is implied). To date, the analogue of such an evolution equation in a medium has not been given. 
It is the object of this Letter to carry out such an extension and to present the first numerical results of such an evolution in 
an extended dense medium.  An alternative approach to this problem leading to a Monte-Carlo implementation 
has been proposed in Ref.~\cite{Armesto:2007dt}.

It should be pointed out that the  
factorized form in Eq.~\eqref{pert_fact} is an assumption that has not yet been proven to hold~\cite{Qiu:1990xy}. So far such a 
factorized form has been shown to hold (at leading twist) in the absence of a medium. Factorization of the 
hard part from the initial distribution has also been demonstrated to hold at all twist in the case of the totally 
inclusive cross section in DIS and up to next-to-leading twist for the single-hadron-inclusive cross section~\cite{Qiu:1990xy}. 
While it may reduce the 
rigor of the present approach, we persist with this form (in the absence of a more accurate formalism), 
seeking justification in 
phenomenological applications. While the LHC and EIC will engender energy scales that are higher by an order of magnitude, 
current experiments at RHIC and HERMES already sample a wide enough range 
of scales to provide a sufficient test for this formalism.

Imagine that a quark jet is produced in a hard collision 
with a large virtuality $M^2$. This virtuality is then reduced by a sequence of partonic emissions. 
The emitted partons will possess a lower virtuality than the parent. These will then decay 
further into lower virtuality branches and the process will continue. The subsequent 
branchings and the development of the partonic shower may be calculated using 
pQCD as long as the virtuality at any vertex is large compared to $\Lambda^2_{QCD}$. Once the 
virtuality drops to some predetermined value $\mu^2$ (a few times $\Lambda^2_{QCD}$), 
a non-perturbative object such as a 
fragmentation function~\cite{Kniehl:2000fe} 
needs to be introduced.  

In order to incorporate the effect of a medium on this cascade process, the fragmentation 
function has to be generalized to include position dependence, i.e.,
\bea
D(z,M^2) \ra D(z,M^2,q^-) |_{\zeta_i^-}^{\zeta_f^-}.  
\eea
$\zeta_i^-$ denotes the location of the production 
of a hard quark with large negative light cone momentum $q^-$  and virtuality $M^2$ 
in an extended dense medium. The leading parton from the shower 
ensuing from such a quark is assumed to exit the 
medium at location $\zeta_f^-$.
We will insist that the final exiting parton which 
fragments to produce the detected hadron possesses virtuality $m^2 (\gg \Lambda_{QCD}^2)$ that is 
much larger than the transverse momentum gained by the jet in traversing the medium (we chose this to be 
$\gtrsim 1$ GeV for jets punching through large nuclei). 

The virtuality drop from $M^2$ to $m^2$ may be achieved through any number of 
emissions. In the case of single emission without scattering, there is no dependence on location and 
the change in $D(z,M^2)$ is given as 
\bea
\mbx \hspace{-0.5cm}
\Delta D_q^h(z,M^2)\!\! &=& \!\!\frac{\A_s}{2\pi}\!\!\!\int\limits_{m^2}^{M^2} \!\!\!\frac{dl_\perp^2}{l_\perp^2} 
\int\limits_z^1 \!\! \frac{dy}{y} P_{q \ra i}(y) D_i^h \left( \frac{z}{y}, m^2 \right) . \label{vacuum_single_emission}
\eea
The splitting function $P_{q \ra i} (y) $ above contains the probability for a quark $q$ to radiate a gluon with momentum fraction $y$ and transverse 
momentum $l_\perp$~\cite{AP}. Virtual corrections which conserve unitarity are implicitly included. 

\begin{figure}[htbp]
\begin{center}
\resizebox{1in}{1in}{\includegraphics[1.in,0.in][5in,4in]{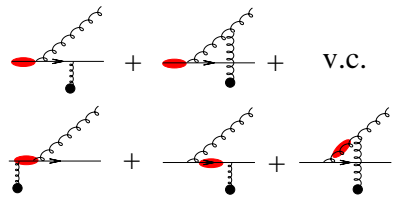}} 
    \caption{ Some of the diagrams included in the quark in-medium splitting function. See text for details.}
    \label{fig4}
\end{center}
\end{figure}

The case of single gluon radiation accompanied by one scattering is given by the diagrams in Fig.~\ref{fig4}.
Only the relevant  amplitudes are drawn. The gluons 
attached to the dark circles are meant to indicate gluons exchanged with the medium when the hard partons scatter off it. 
The light red (shaded) ellipses on the Feynman diagrams 
denote the propagators that are considerably more virtual than the remaining 
parton lines. To the five 
diagrams included, one must add contributions with no scattering as well as contributions
denoting the interference of double scattering with no scattering diagrams and virtual diagrams (denoted as \emph{v.c.}). 
The two diagrams in the top row of Fig.~\ref{fig4}, 
where the initial parton with a large virtuality splits into two partons with lower virtuality, followed by a 
single soft scattering, is similar to the case in vacuum.
The single soft scattering merely introduces a mild change in direction of propagation of the struck line. 
These diagrams are referred to as the vacuum-like contribution to in-medium induced 
radiation and their interference with the 
remaining diagrams as the vacuum-medium-induced interference contribution. The  
three diagrams in the bottom row represent the case where the scattering with a hard gluon in the medium 
leads to a rise in the virtuality, immediately lost by gluon emission~\cite{guowang}. 

The result of the interference of all such diagrams [which leads to the Landau-Pomeranchuck-Migdal (LPM) effect] 
yields the ``medium dependent'' correction to the in-medium 
modified fragmentation function. If the drop in virtuality from $M$ to $m$ were to occur solely through 
single scattering and single emission, then~\cite{Majumder:2004pt,guowang}, 
\bea
\Delta {D_q^h}^1(z,M^2,q^-)\lvert_{\zeta_i}^{\zeta_f} \!\!\!&=&\!\!\! \frac{\A_s}{2\pi}\! \int\limits_{m^2}^{M^2} \!\frac{dl_\perp^2}{l_\perp^2} 
\!\int\limits_z^1 \! \frac{dy}{y} \frac{\tilde{P}_{q \ra i}(y)}{l_\perp^2} 
\int\limits_{\zeta_i}^{\zeta_f} d \zeta   \nn \\
\ata\!\!\!\!\frac{2\pi \A_s }{N_c}\! \left[ 2 - 2 \cos\left\{  \frac{l_\perp^2 ( \zeta - \zeta_i) }{2q^- y(1-y)}\right\} \right]  
 \nn \\
\ata \!\!\!\!\rho_g(\zeta,x_T)D_i^h \!\left( \frac{z}{y}, m^2 \right) . \label{medium_single_emission} 
\eea  
In the equation above, $\rho_g(\zeta)$ is the gluon density at $\zeta$. 
The ($-$) superscript on the locations have been dropped for simplicity. 
The tilde on the splitting function indicates that the color factor and virtual correction are somewhat different from the 
case of the vacuum (requiring the final outgoing line, after a scattering and an emission to be nearly on-shell 
restricts the range of $y$, see Ref.~\cite{guowang} for details).  In Eq.~\eqref{medium_single_emission},
$\zeta$ represents the 
location of the scattering vertex of the hard parton off the medium in Fig.~\ref{fig4}.
Due to interference effects, the medium modified 
fragmentation function is now a function of the jet energy $q^-$; it is no longer universal and 
strongly depends on the details of the medium as introduced by the density 
factor $\rho_g(\zeta)$ and its space-time distribution. However, the functional 
dependence of the fragmentation function on the medium density is still universal and thus 
may be used to compute the medium modified fragmentation function in any medium where 
$\rho_g(\zeta)$ is known.

Given that each radiation leads to a drop in the virtuality of the propagating 
parton, successive radiations are assumed to be strongly ordered in transverse momentum. While this is well known for the
case of radiations in the vacuum, 
in a medium, the strong ordering of the virtualities will be broken if the parton 
encounters a hard scattering which will lead to a large transverse 
momentum radiation. 
At next-to-leading twist, as is the case in this calculation, 
this hard scattering can at most occur at the location of the first rescattering (in the case of higher twist contributions as in Ref.~\cite{Majumder:2007ne}, this hard 
scattering may occur at any later rescattering). 
In such an ordered scenario, we may write down the expression for the parton to lose virtuality from 
$M^2$ to $m^2$ in two ordered emissions as (we suppress $q^-$ in the argument) 
\bea
\Delta {D_q^h}^2(z,M^2)\lvert_{\zeta_i}^{\zeta_f} \!\!\!&=&\!\!\! \frac{\A_s}{2\pi}\! 
\int\limits_{m^2}^{M^2} \!\frac{dl_\perp^2}{l_\perp^2} 
\!\int\limits_z^1 \! \frac{dy}{y} \frac{\tilde{P}_{q \ra i}(y)}{l_\perp^2} 
\int\limits_{\zeta_i}^{\zeta_f} d \zeta  \nn \\
\ata \frac{2\pi \A_s }{N_c}  \left[ 2 - 2 \cos\left\{  \frac{l_\perp^2 ( \zeta - \zeta_i) }{2q^- y(1-y)}\right\} \right]  
 \nn \\
\!\!\!\ata\!\!\!\rho_g (\zeta) \frac{\A_s}{2\pi}\! \int\limits_{m^2}^{l_\perp^2} \!\frac{d {l_1}_\perp^2}{{l_1}_\perp^2} 
\!\!\int\limits_{z/y}^1 \!\! \frac{dy_1}{y_1} \frac{\tilde{P}_{i \ra j}(y_1)}{{l_1}_\perp^2} 
\!\!\int\limits_{\zeta}^{\zeta_f} \!d \zeta_1\!  \nn \\
\ata \!\!\!\frac{2\pi \A_s }{N_c}\! \left[ 2 - 2\! \cos\!\left\{\!  \frac{l_\perp^2 ( \zeta_1 - \zeta) }{2q^- y y_1(1-y_1)}\!\right\}\! \right]  
 \nn \\
\ata \rho_g (\zeta_1) D_j^h \!\left( \frac{z}{yy_1}, m^2 \right) . \label{medium_double_emission} 
\eea
Due to the drop in virtuality after the first emission, the location of the first 
scattering in the medium (whether soft or hard), is considered as the 
origin of the $\zeta_1$ integration and thus of the interference pattern connected 
with the second scattering. Alternatively stated, this means that 
the above expression 
focuses only on 
ladder diagrams both in the scattering and in the emitted gluon sector. 
The insistence on this is simply based 
on the dominance of terms which contain a strong ordering of virtualities in successive emissions. 

In the strongly ordered $l_\perp$, or planar emission limit, the emission points for successive lower $l_\perp$ emissions 
are also strongly ordered.
The extension to multiple emissions, may now be written down similar to Eq.~\eqref{medium_double_emission}. 
Differentiating the equation for $D(z,M^2)|_{\zeta_i}^{\zeta_f}$ (containing multiple emissions) with respect to $\log(M^2)$ 
leads to the ``medium 
dependent part'' of the evolution equations for the medium modified fragmentation function,
\bea
\frac{\prt {D_q^h}(z,M^2\!\!,q^-)|_{\zeta_i}^{\zeta_f}}{\prt \log(M^2)} \!\!\!&=& \frac{\A_s}{2\pi} \int\limits_z^1 \frac{dy}{y} 
\int\limits_{\zeta_i}^{\zeta_f} d\zeta \tilde{P}(y) K_{q^-,M^2} ( y,\zeta) \nn \\ 
\ata  {D_q^h}\left. \left(\frac{z}{y},M^2\!\!,q^-y\right) \right|_{\zeta}^{\zeta_f}.  \label{in_medium_evol_eqn}
\eea
The single scattering kernel $K$ is given as 
\bea
K = \frac{2\pi \A_s \rho(\zeta)}{N_c} 
\left[ 2 - 2 \cos\left\{ \frac{M^2 (\zeta - \zeta_i)}{ 2 q^- y (1- y)} \right\}  \right] .
\eea
The full evolution equation for the medium modified fragmentation functions will include
contributions from pure vacuum splitting functions as well as contributions from gluon fragmentation 
functions which have a similar in-medium evolution. 

\begin{figure}[htbp]
\resizebox{1.8in}{2.25in}{\includegraphics[1in,0.6in][4in,4.6in]{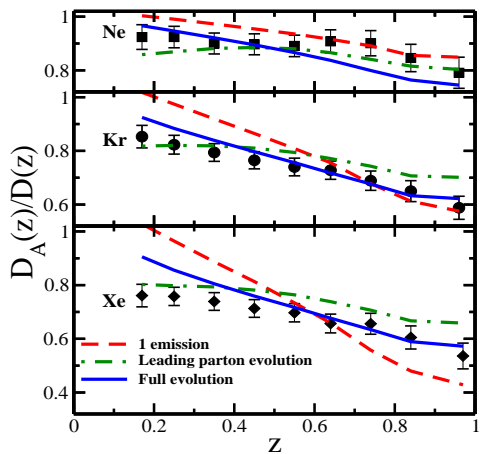}} 
    \caption{ Comparison of the experimental results on the nuclear attenuation observed in the DIS 
on $Ne$, $Kr$ and $Xe$ with three different calculations of the ratio of the medium modified to the vacuum 
fragmentation functions. See text for details. The data points (and the bins of $\nu$ and $Q^2$) are from Ref.~\cite{Airapetian:2007vu}. } 
    \label{fig7}
\end{figure}

The simplest application of the formalism developed above is to compute 
the medium modified fragmentation function in the case of DIS on a large nucleus,
where, at least, one hadron with a large forward momentum is detected in the final state. 
Experiments present the ratio of
this with the vacuum fragmentation function at the same $z$ and $Q^2$, called the nuclear attenuation factor~\cite{Airapetian:2000ks}.
Experimental results for three nuclei ($Ne$, $Kr$ and $Xe$) are presented in Fig.~\ref{fig7}.
Assuming single scattering and single in-medium emission, the medium modified fragmentation function may be calculated using 
Eq.~\eqref{medium_single_emission}; the corresponding attenuation factor is the red dashed line in Fig.~\ref{fig7}. 
In the case of multiple emissions, one will need to 
use the in-medium evolution equations of Eq.~\eqref{in_medium_evol_eqn}. In either case, the basic kernel 
$K_{q^-,M^2} (y,\zeta)$ is identical. To calculate this kernel, we introduce a nucleon density in the large nucleus as
 an input Ansatz. Due to the simplicity in analytic calculations, 
we use a hard sphere density distribution (with radius $R_A$): $\rho(\zeta) = \rho_0 \theta (R_A - |\zeta| )$.
While $R_A$ depends on the nucleus, $\rho_0$ is a fit parameter dialed to obtain the best overall fit. 
We also approximate $D(z)|_\zeta^{\zeta_f} \simeq D(z)|_{\zeta_i}^{\zeta_f}$, which greatly speeds calculation.

As may be seen from the comparison with the data in Fig.~\ref{fig7}, the fit, for the case of single 
scattering and single emission (red dashed lines), with smaller nuclei such as 
$Ne$ is adequate. The comparison, progressively worsens as one proceeds to larger 
nuclei. This is clearly seen in the case for $Xe$ where there seems to be almost a different 
slope with $z$ between the calculations and the experimental results. In some ways, this is to be 
expected; as one proceeds to larger nuclei, the possibility of multiple scattering and multiple 
emission increases and the results of a formalism which only included single scattering and 
single emission in medium should show a systematic departure with nuclear size. 
This is to be contrasted with the medium modified 
fragmentation functions calculated using Eq.~\eqref{in_medium_evol_eqn} (solid blue lines). 
One immediately sees a marked improvement in the 
comparison between the calculations and the experimental data. This represents the principal result of 
this Letter.

While the inclusion of multiple emissions, each with a single scattering kernel leads to a marked improvement, 
there still is a small difference between the data points and the theoretical curves for the case of 
DIS on $Xe$ at small $z$. One way to improve this is to include multiple scatterings per emission as was 
done for photon emission in Ref.~\cite{Majumder:2007ne}. Such an effect is expected to become important for larger $A$,  
but will have to be left for a future effort. 
Yet another effect is the presence of a certain amount of hadronic energy loss~\cite{Accardi:2007in}, in addition to the partonic energy loss calculated in this Letter. 
This effect should be most dominant at lower $z$. To estimate its importance, we introduce an phenomenological 
extension as indicated by the green dot-dashed line in Fig.~\ref{fig7}.
In this case the contribution from the gluon fragmentation functions is removed, i.e., hadrons emanating from the fragmentation of 
soft gluons radiated by the jet are completely stopped in
the medium. This latter calculation has the best agreement with the largest nucleus ($Xe$), where the 
agreement with data is remarkable. The actual attenuation will lie in between the green dot-dashed and the blue 
solid lines. In each of the calculations, the gluon density was dialed to provide the best overall fit to all three data sets. 
This leads to a small spread in the medium gluon densities ($\rho_0$) required, which corresponds to a spread in the quark jet quenching parameter $\hat{q}_0$ (see Ref.~\cite{Majumder:2007ae}) between 0.06 to 0.18 GeV$^2/$fm.

Applications of the methods presented in this Letter to the case of high transverse momentum hadron production in heavy-ion collisions 
has recently been presented in Ref.~\cite{Bass:2008rv}, where Eq.~\eqref{in_medium_evol_eqn} was evaluated in a three dimensional 
hydro-dynamically expanding medium. Due to the larger error bars in the data at RHIC, the improvement of the present formalism over 
the medium modified fragmentation functions calculated in the single emission limit (in an identical medium, using Eq.\eqref{medium_single_emission}~\cite{Majumder:2007ae}) are somewhat less evident.
 
%
This work was supported by the U.S. Department of Energy
under grant nos. DE-FG02-05ER41367 and DE-FG02-01ER41190. 
The author thanks B.~M\"{u}ller U.~Heinz, G.-Y.~Qin and X.~N.~Wang for discussions.

\end{document}